\documentstyle[12pt,rotate,worldsci]{article}
\input{psfig}

\newcommand{\rf}[4]{{\em {#1}} {\bf #2}, #3 (#4)}

\newcommand{\pr}{Phys.\ Rev.\ }

\newcommand{\pl}{Phys.\ Lett.\ }

\newcommand{\np}{Nucl.\ Phys.\ }

\def\be{\begin{equation}}
\def\ee{\end{equation}}
\def\bea{\begin{eqnarray}}
\def\eea{\end{eqnarray}}
\def\Tr{{\rm Tr}\,}
\def\muhat{\hat{\mu}}
\def\qhat{\hat{q}}
\def\bra{\langle}
\def\ket{\rangle}
\def\half{\frac{1}{2}}
\def\sss{\scriptscriptstyle}
\def\Muv{M_{\sss\rm UV}}
\def\Mir{M_{\sss\rm IR}}

\newcommand{\err}[2]{\raisebox{-0.4ex}
{$\stackrel{\scriptstyle +#1}{\scriptstyle -#2}$}}

\newcommand{\bb}{\bibitem}

\begin{document}
\title{LATTICE CALCULATION OF THE GLUON PROPAGATOR AND\\
THE STRANGENESS MAGNETIC MOMENT OF THE NUCLEON}
\author{Anthony G.\ Williams\\
{\em CSSM and Department of Physics and Mathematical Physics\\
University of Adelaide, Australia 5005\\
{\small \tt awilliam@physics.adelaide.edu.au\\ 
     http://www.physics.adelaide.edu.au/cssm}
    }}

\maketitle
\setlength{\baselineskip}{2.6ex}

\vspace{0.7cm}
\begin{abstract}

This contribution summarizes a
presentation combining two topics in lattice QCD.
Firstly, the gluon propagator in Landau gauge is
calculated in quenched QCD on a large ($32^3\times 64$) lattice at
$\beta=6.0$.  New structure
seen in the infrared region survives conservative cuts to the lattice
data, and serves to exclude a number of models that have appeared in
the literature.  Secondly, I report 
on a recent lattice QCD calculation
of the strangeness magnetic moment of the nucleon. The result is
$G_M^s(0) =  - 0.36 \pm 0.20 $. The strangeness Sachs electric mean-square
radius $\langle r_s^2\rangle_E$ was also found to be small and negative.

\end{abstract}
\vspace{0.7cm}

\section{Introduction}

Here I summarize the central results described in a single long
parallel session talk, which was comprised of two distinct
studies of lattice QCD.  The first of these is concerned with the
direct lattice calculation of the Landua gauge nonperturbative gluon
propagator, an understanding of which is central to our understanding
of the nature of confinement.  The second topic is a lattice calculation
of the strangeness magnetic moment of the nucleon.  Different
confining quark models predict a variety of results for this quantity
and so lattice calculations provide a benchmark
against which we can test our understanding of hadron structure.

\section{Lattice Calculation of the Gluon Propagator}

\subsection{Motivation}

The infrared behaviour of the gluon propagator is important for an
understanding of confinement.  Previous conjectures range from a
strong divergence~\cite{mandelstam,bp} to a propagator that vanishes
in the infrared~\cite{gribov,stingl}.
Lattice QCD should in principle be able to resolve this issue by
first-principles, model-independent calculations.  However, lattice
studies have been inconclusive up to now,\cite{bps,mms}
since they have not been able to access sufficiently low momenta.  The
lower limit of the available momenta on the lattice is given by
$q_{\rm min} = 2\pi/L$, where $L$ is the length of the lattice.
Here we will report results using a lattice with a length of 3.3~fm in
the spatial directions and 6.7~fm in the time direction.  This gives us
access to momenta as small as 400~MeV.

\subsection{Summary of Formalism}
\label{sec:def}

The gluon field $A_\mu$ can be extracted from the link variables
$U_\mu(x)$ using
$U_\mu(x) =  e^{ig_0aA_\mu(x+\muhat/2)} + {\cal O}(a^3)$.
Inverting and Fourier transforming this, we obtain
\bea
A_\mu(\qhat) & \equiv & \sum_x e^{-i\qhat\cdot(x+\muhat/2)}
 A_\mu(x+\muhat/2) \nonumber \\
 & = & \frac{e^{-i\qhat_{\mu}a/2}}{2ig_0a}
       \left[\left(U_\mu(\qhat)-U^{\dagger}_\mu(-\qhat)\right)
 - \frac{1}{3}\Tr\left(U_\mu(\qhat)-U^{\dagger}_\mu(-\qhat)\right)\right] , 
\eea
where $U_\mu(\qhat)\equiv\sum_x e^{-i\qhat x}U_\mu(x)$ and
$A_\mu(\qhat)\equiv t^a A_{\mu}^a(\qhat)$.  The available momentum
values $\qhat$ are given by
$\qhat_\mu  = 2 \pi n_\mu/(a N_\mu)$, $\ n_\mu=0,\ldots,N_\mu-1$,
where $N_\mu$ is the number of points in the $\mu$ direction.  The
gluon propagator $D^{ab}_{\mu\nu}(\qhat)$ is defined as
$D^{ab}_{\mu\nu}(\qhat) = \bra A^a_\mu(\qhat)
A^b_\nu(-\qhat) \ket\,/\,V$.
The Landau gauge propagator in the continuum has the
structure
$D_{\mu\nu}^{ab}(q) =
\delta^{ab}[\delta_{\mu\nu}-(q_{\mu}q_{\nu}/q^2)]D(q^2)$.
At tree level, $D(q^2)$ will have the form $D^{(0)}(q^2) = 1/q^2$.
On the lattice, this becomes
\be
D^{(0)}(\qhat) = 
1/\sum_{\mu}\left(\frac{2}{a}\sin\frac{\qhat_{\mu}a}{2}\right)^2\, .
\label{eq:lat-tree}
\ee
Since QCD is asymptotically free, we expect that up to logarithmic
corrections, $q^2 D(q^2) \to 1$ in the ultraviolet.  Hence we define
the new momentum variable $q$ by
$q_\mu \equiv (2/a)\sin(\qhat_\mu a/2)$
and work with this throughout.
The (bare) lattice gluon propagator is related to the renormalised
continuum propagator $D_R(q;\mu)$ via
$D^L(qa) = Z_3(\mu,a) D_R(q;\mu)$.
The renormalisation constant $Z_3(\mu,a)$ can be found by imposing a
momentum subtraction renormalisation condition
$D_R(q)|_{q^2=\mu^2} = 1/\mu^2$.
The asymptotic behaviour of the renormalised gluon propagator in the
continuum is given to one-loop level by
$D_R(q) \equiv D_{\rm bare}(q)/Z_3
=(1/q^2)\left[(1/2)\ln(q^2/\Lambda^2)\right]^{-d_D}$
with 
$d_D=(39-\xi-4N_f)/[4(33-2N_f)]=13/44$,
where both the gauge parameter $\xi$ and
the number of fermion flavours $N_f$ are zero in this calculation.

\subsection{Simulation Parameters, Finite Size Effects and Anisotropies}
\label{sec:params}

We have analysed three lattices, with different values for the volume
and lattice spacing.  The details are given in Table~\ref{tab:sim-params}.
In the following, we are particularly interested in the deviation of
the gluon propagator from the tree level form.  We will therefore
factor out the tree level behaviour and plot $q^2 D(q^2)$ rather than
$D(q^2)$ itself.

\begin{table}[tb]
\begin{center}
\leavevmode
\begin{tabular}{lcccrcc}
\hline
Name &$\beta$ &$a_{\rm st}^{-1}$ (GeV) &Volume &$N_{\rm conf}$ &$a \widehat
q_{\rm Max}$ & $|\partial_\mu A_\mu|$ \\ 
\hline
\hline
Small    &6.0  &1.885   &$16^3\times 48$ &125   &$2 \pi / 4$ & $< 10^{-6}$ \\
Large    &6.0  &1.885   &$32^3\times 64$ &75    &$2 \pi / 4$ & $< 10^{-6}$ \\
Fine     &6.2  &2.63    &$24^3\times 48$ &223   &$2 \pi / 4$ & $< 10^{-6}$ \\
\hline
\end{tabular}
\end{center}
\caption{Simulation parameters}
\label{tab:sim-params}
\end{table}

\begin{figure}[bt]
\begin{center}
\leavevmode
\mbox{\rotate[l]{\psfig{figure=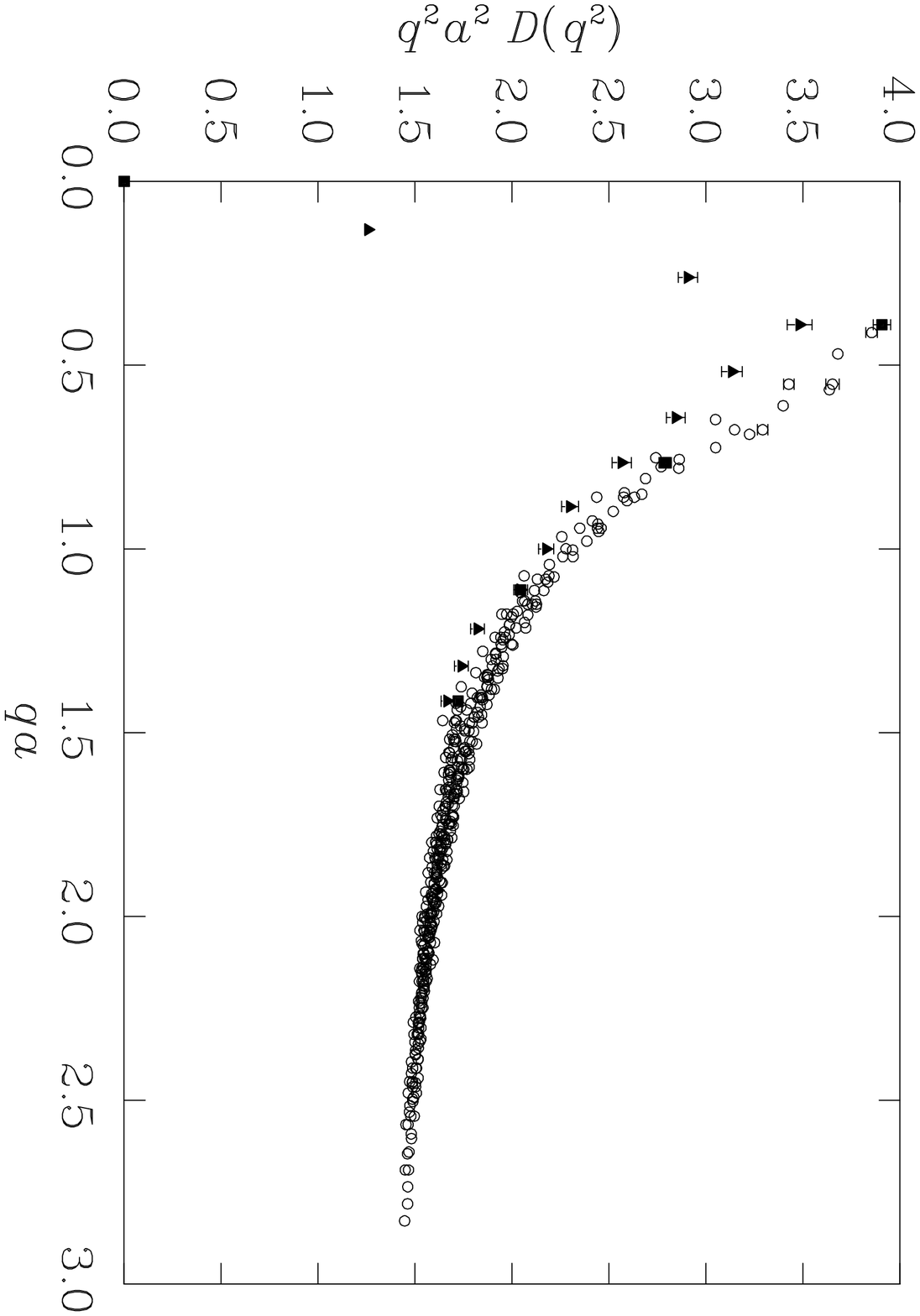,height=2.2in}}\hspace{0.5cm}
\rotate[l]{\psfig{figure=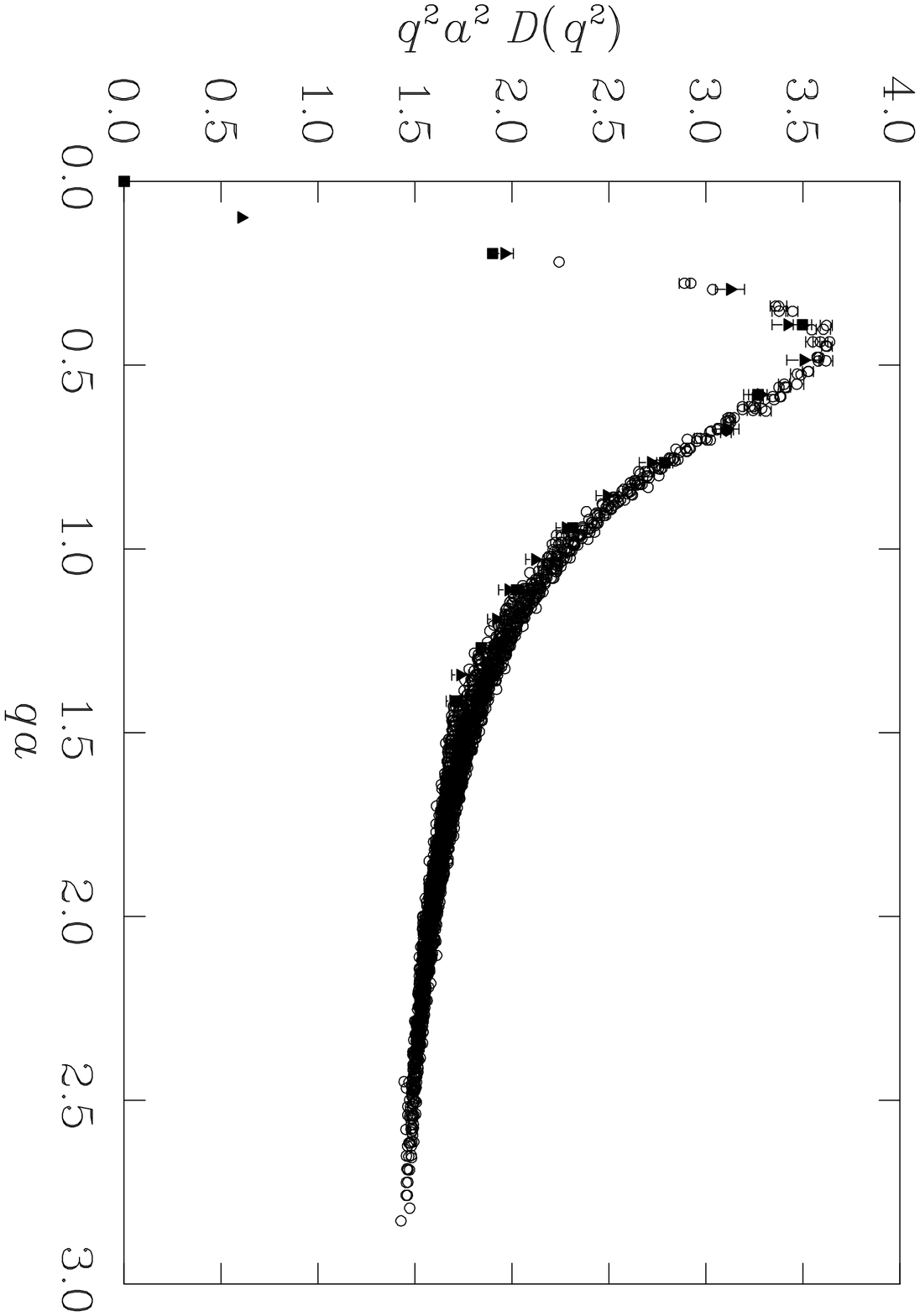,height=2.2in}}}
\end{center}
\caption{Componentwise data, for the small lattice (left) and the
large lattice (right).  The filled triangles denote momenta directed
along the time axis, while the filled squares denote momenta directed
along one of the spatial axes.}
\label{fig:cpt-data}
\end{figure}

Fig.~\ref{fig:cpt-data} shows the gluon propagator as a function of
$qa$ for the small and large lattices, with momenta in different
directions plotted separately.  For low momentum values on the small
lattice, there are large discrepancies due to finite size effects
between points representing momenta along the time axis and those
representing momenta along the spatial axes.  These discrepancies are
absent from the data from the large lattice, indicating that finite
size effects here are under control.

However, at higher momenta, there are anisotropies which remain for
the large lattice data, and which are of approximately the same
magnitude for the two lattices.  In order to eliminate these
anisotropies, which arise from large momenta (i.e., finite lattice spacing
errors), we select momenta lying within a cylinder of radius
$\Delta\qhat a = 2\times 2\pi/32$ 
along the 4-dimensional diagonals in order to distribute
large momenta across the four momentum components.\cite{letter}

\subsection{Scaling behaviour}

\begin{figure}[t]
\begin{center}
\leavevmode
\mbox{\rotate[l]{\psfig{figure=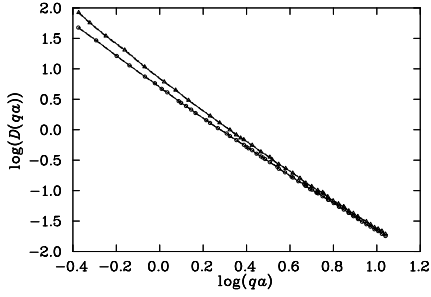,height=2.2in}}}
\end{center}
\caption{The dimensionless, unrenormalised gluon propagator as a
function of $\ln(qa)$ for the two values of $\beta$.  The triangles
denote the data for the small lattice at $\beta=6.0$, while the
circles denote the data for $\beta=6.2$.}
\label{fig:compare_data_lattmom}
\end{figure}

Since the renormalised propagator $D_R(q;\mu)$ is independent of the
lattice spacing, we can derive a simple,
$q$-independent expression for the ratio of the unrenormalised
lattice gluon propagators at the same value of $q$:
\be
\frac{D_c(qa_c)}{D_f(qa_f)} = 
\frac{Z_3(\mu,a_c)D_R(q;\mu)/a_c^2}{Z_3(\mu,a_f)D_R(q;\mu)/a_f^2}
= \frac{Z_c}{Z_f}\frac{a_f^2}{a_c^2} \, ,
\label{eq:gluon_match_ratio}
\ee
where the subscript $f$ denotes the finer lattice ($\beta=6.2$ in this
study) and the subscript $c$ denotes the coarser lattice
($\beta=6.0$).  We can use this relation to study directly the scaling
properties of the lattice gluon propagator by matching the data for
the two values of $\beta$.  This matching can be performed by
adjusting the values for the ratios $R_Z = Z_f/Z_c$ and $R_a =
a_f/a_c$ until the two sets of data lie on the same curve.
Fig.~\ref{fig:compare_data_lattmom} shows the data for both lattice
spacings as a function of $qa$ before shifting.  
This procedure gives
us an estimate of $R_a=0.745\err{32}{37}$, which is fully compatible
with the string-tension value~\cite{bs} of $0.716\pm 0.040$.  The
corresponding estimate for the ratio of the renormalisation constants is
$R_Z = 1.038\err{26}{21}$.

\subsection{Model Fits}
\label{sec:models}

We have demonstrated scaling in our lattice data over the entire range
of $q^2$ considered, and will now proceed with model fits.
The following functional forms have been considered:
\bea
D(q^2) = & 
\frac{Z q^2}{q^4+M^4}\left(\half\ln\frac{q^2+M^2}{M^2}\right)^{-d_D}
 & \hspace{-0.3cm}\mbox{(Gribov\,\cite{gribov})}
\label{model-first}
\label{model:lita}
\label{model:gribov} \\
D(q^2) = & Z\frac{q^2}{q^4+2A^2q^2+M^4}
\left(\half\ln\frac{q^2+M^2}{M^2}\right)^{-d_D}
 & \hspace{-0.1cm}\mbox{\hfill(Stingl\,\cite{stingl})}
\label{model:stingl} \\
D(q^2) = & \frac{Z}{(q^2)^{1+\alpha}+M^2}
 & \hspace{-1.7cm}\mbox{\hfill(Marenzoni {\em et al}\,\,\cite{mms})}
\label{model:marenzoni} \\
D(q^2) = & Z
\left[(q^2+M^2(q^2))\ln\frac{q^2+4M^2(q^2)}{\Lambda^2}\right]^{-1}
 & \hspace{-0.6cm}\mbox{(Cornwall\,\cite{cornwall})}
\label{model:cornwall} \\
 \mbox{where} & 
M(q^2) = M\left\{\ln\frac{q^2+4M^2}{\Lambda^2}/
\ln\frac{4M^2}{\Lambda^2}\right\}^{-6/11} \nonumber \\
D(q^2) = & Z\left(\frac{A}{(q^2+\Mir^2)^{1+\alpha}} + 
\frac{1}{q^2+\Muv^2}\left(\half\ln\frac{q^2+\Muv^2}{\Muv^2}\right)^{-d_D}
 \right)
\label{modelA} \\
D(q^2) = & Z\left(\frac{A}{(q^2)^{1+\alpha}+(\Mir^2)^{1+\alpha}} +
\frac{1}{q^2+\Muv^2}\left(\half\ln\frac{q^2+\Muv^2}{\Muv^2}\right)^{-d_D}
 \right)
\label{modelB} \\
D(q^2) = & Z\left(A e^{-(q^2/\Mir^2)^{\alpha}} + 
\frac{1}{q^2+\Muv^2}\left(\half\ln\frac{q^2+\Muv^2}{\Muv^2}\right)^{-d_D}
   \right)
\label{modelC}
\label{model-last}
\eea

\begin{figure}[tb]
\begin{center}
\leavevmode
\mbox{\rotate[l]{\psfig{figure=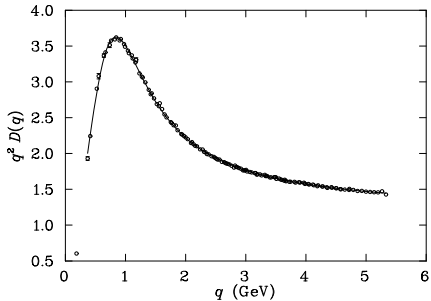,height=2.2in}}}
\end{center}
\caption{The gluon propagator multiplied by $q^2$, with nearby points
averaged.  The line illustrates our best fit to the form defined in
(\protect\ref{modelB}).  The fit is performed over all points shown,
excluding the one at the lowest momentum value, which may be sensitive
to the finite volume of the lattice.  The scale is taken from the
string tension~\protect\cite{bs}.  }
\label{fig:fit-modelB}
\end{figure}

We have also considered~\cite{next} special cases of the three forms
(\ref{modelA})--(\ref{modelC}), with $\Mir=\Muv$ or with specific
values for the exponent $\alpha$.  Equations (\ref{model:gribov}) and
(\ref{model:stingl}) are modified in order to
exhibit the appropriate one-loop asymptotic behaviour.
Models~(\ref{modelA}) and (\ref{modelB})
are constructed as generalisations of (\ref{model:marenzoni}) with the
correct dimension and asymptotic behaviour.
All models were fitted to the large lattice data using the cylindrical
cut.  The lowest momentum value was excluded, as the volume dependence
of this point could not be assessed.  In order to balance the
sensitivity of the fit between the high- and low-momentum region,
nearby data points within $\Delta(qa) < 0.05$ were averaged.
$\chi^2$ per degree of freedom and parameter values for fits to all
these models are shown in Table~\ref{tab:fit-params}.  It is clear
that model (\ref{modelB}) accounts for the data better than any of the
other models.  The best fit to this model is illustrated in
Fig.~\ref{fig:fit-modelB}.

\begin{table}
\begin{tabular}{c|c|ccccc}
Model & $\chi^2/$dof & $Z$ & $A$ & $\Mir$ & $\Muv$ or $\Lambda$ & $\alpha$
\\ \hline
\ref{model:gribov} & 1972 & 2.19\err{31}{15} & & 0.23\err{14}{1} \\
\ref{model:stingl} & 1998 & 2.2 & 0 & 0.23 \\
\ref{model:marenzoni} & 163 & 2.41\err{0}{12} & & 0.14\err{4}{14}
 & & 0.29\err{6}{2} \\
\ref{model:cornwall} & 50.3 & 6.5\err{7}{9} & & 0.24\err{3}{16} &
0.27\err{7}{7}  \\
\ref{modelA} & 2.96 & 1.54\err{10}{20} & 1.24\err{21}{21}
 & 0.46\err{2}{14} & 0.96\err{47}{17} & 1.31\err{16}{43} \\
\ref{modelA}; $\Mir\!=\!\Muv$ & 3.73 & 1.71\err{9}{0} & 0.84\err{0}{29} & 0.48
\err{2}{17}
 & & 1.52\err{12}{37} \\
\ref{modelB} & 1.57 & 1.78\err{45}{20} & 0.49\err{17}{6}
 & 0.43\err{5}{1} & 0.20\err{37}{19} & 0.95\err{7}{5} \\
\ref{modelB}; $\Mir\!=\!\Muv$ & 4.00 & 1.62\err{3}{4} & 0.58\err{5}{1} & 0.40
\err{9}{2}
 & & 0.92\err{17}{1} 
\\
\ref{modelC} & 47 & 2.09\err{30}{12} & 29\err{166}{2}
 & 0.22\err{0}{16} & 0.14\err{11}{10} & 0.49\err{0}{16}
\end{tabular}
\caption{Parameter values for fits to models
(\protect\ref{model-first})--(\protect\ref{model-last}).  The
values quoted are for fits to the entire set of data.  The errors
denote the uncertainties in the last digit(s) of the parameter values
which result from varying the fitting range.  The fitting ranges
considered when evaluating the uncertainties are those with a minimum
of 40 points included and with the minimum value for $qa$ no larger
than 1.03 (point number 40).}
\label{tab:fit-params}
\end{table}

\subsection{Summary of Numerical Results}
\label{sec:discuss}

We have evaluated the gluon propagator on an asymmetric lattice with a
large physical volume.  By studying the anisotropies in the data, and
comparing the data with those from a smaller lattice, we have been
able to conclude that finite size effects are under control on the
large lattice.
A clear turnover in the behaviour of $q^2 D(q^2)$ has been observed at
$q \sim 1$GeV, indicating that the gluon propagator diverges less
rapidly than $1/q^2$ in the infrared, and may be infrared finite or
vanishing. 
The data are consistent with a functional form $D(q^2) = D_{\rm IR} +
D_{\rm UV}$, where
\be
D_{\rm IR} = \half \frac{1}{q^4+M^4} ,
\label{eq:ir-prop}
\ee
$M \sim 1$ GeV, and $D_{\rm UV}$ is the appropriate asymptotic form.
A more detailed analysis~\cite{next} of the
asymptotic behaviour reveals that the one-loop formula
remains insufficient at $q^2=50{\rm GeV}^2$.
Issues for future study include
the effect of Gribov copies and of dynamical fermions.
We also hope to use improved actions to
perform realistic simulations at larger lattice spacings.  This would
enable us to evaluate the gluon propagator on larger physical volumes,
giving access to lower momentum values.

\section{Nucleon Strangeness Magnetic Moment}

\subsection{Motivation}

We summarize some recent results \cite{dlw} on nucleon electromagentic form
factors, including the strangeness electric and magnetic form factors.
The strangeness content of the nucleon has been a topic of considerable
recent interest for a variety of reasons. The studies of nucleon
spin structure functions in polarized deep inelastic
scattering experiments at CERN and SLAC \cite{DIS}, combined with
neutron and hyperon $\beta$ decays, have turned up a surprisingly
large and negative polarization from the strange quark. In addition,
there is a well-known long-standing discrepancy between the
pion-nucleon sigma term
extracted from the low energy pion-nucleon scattering~\cite{gls91}
and that from the octect baryon masses~\cite{cheng76}. This
discrepancy can be reconciled if a significant $\bar{s}s$ content
in the nucleon~\cite{cheng76,gl82} is admitted.
To address some of these issues, an experiment to
measure the neutral weak magnetic form factor $G^Z_M$
via elastic parity-violating electron scattering 
at backward angles was recently carried out
by the SAMPLE collaboration~\cite{SAMPLE97}. The strangeness magnetic form
factor is obtained by subtracting out the nucleon magnetic form
factors $G^p_M$ and $G^n_M$. The reported value is
$G^s_M(Q^2=0.1$GeV$^2)= +0.23\pm 0.37\pm 0.15\pm 0.19 $ and does not
yet provide a strong constraint on the sign.

Theoretical predictions of $G^s_M(0)$ vary widely. The values
from various models and analyses range from $ -0.75\pm 0.30 $
in a QCD equalities analysis~\cite{lei96} to $+ 0.37 $ in an $SU(3)$
chiral bag model~\cite{hpm97}. While a few give positive
values~\cite{hpm97,gi97}, most model predictions are negative with a
typical range of $-0.25$ to 
$-0.45$. Summaries of these predictions can be found in
Refs.~\cite{lei96,cbk96}  A similar situation exists for the
strangeness electric mean-square radius $\langle r_s^2\rangle_E$.
A number of the predictions are positive while a few
are negative.
Elastic $\vec{e}\; p$ and $\vec{e}\;{}^{4}He$ parity-violation experiments
are currently planned at TJNAF~\cite{pve91} to
measure the asymmetry $A_{LR} $ at forward angles to extract
$\langle r_s^2\rangle_E$. Hopefully, they will settle
the issue of its sign.

\subsection{Numerical Results}

The lattice formulation of the electromagnetic and other form factors has
been given
in detail in the past~\cite{dll9596,dwl90}. Here, we shall concentrate on the
DI contribution, where the strangeness current contributes.
In the Euclidean formulation, the Sachs EM form factors
can be obtained by the combination of two- and three-point functions
\begin{equation} \label{twopt}
G_{NN}^{\alpha\alpha}(t,\vec{p}) = \sum_{\vec{x}}e^{-i\vec{p}\cdot
\vec{x}}  \langle 0| \chi^\alpha(x) \bar{\chi}^\alpha(0) |0 \rangle
\end{equation}
\begin{equation} \label{threept}
G_{NV_{\mu}N}^{\alpha\beta}(t_f,\vec{p},t,\vec{q})=
\sum_{\vec{x}_f,\vec{x}} e^{-i\vec{p}\cdot\vec{x}_f
  +i\vec{q}\cdot\vec{x}} \langle 0| \chi^\alpha(x_f) V_\mu(x)
  \bar{\chi}^\beta(0) |0 \rangle ,
\end{equation}
where $\chi^\alpha$ is the nucleon interpolating field and $V_{\mu}(x)$
the vector current.
With large Euclidean time separation, i.e. $t_f - t >> a$ and $t >> a$,
where $a$ is the lattice spacing,
\begin{eqnarray}
\frac{\Gamma_i^{\beta\alpha}G_{NV_jN}^{\alpha\beta}(t_f,\vec{0},t,
\vec{q})} {G_{NN}^{\alpha\alpha}(t_f,\vec{0})}
\frac{G_{NN}^{\alpha\alpha}(t,\vec{0})}{G_{NN}^{\alpha\alpha}
(t,\vec{q})} \longrightarrow  \frac{\varepsilon_{ijk}q_k}{E_q + m}
 G_M(q^2),  \label{mff} \\
\frac{\Gamma_E^{\beta\alpha}G_{NV_4N}^{\alpha\beta}(t_f,\vec{0},t,
\vec{q})} {G_{NN}^{\alpha\alpha}(t_f,\vec{0})}
\frac{G_{NN}^{\alpha\alpha}(t,\vec{0})}{G_{NN}^{\alpha\alpha}
(t,\vec{q})} \longrightarrow   G_E(q^2), \label{eff}
\end{eqnarray}
where $\Gamma_i= \sigma_i (1 + \gamma_4)/2$ and 
$\Gamma_E = (1 + \gamma_4)/2$.

We shall use the conserved current from the Wilson action which, being
point-split, yields slight variations
on the above forms and these are given in Ref.~\cite{dwl90}
Our 50 quenched gauge configurations were generated on a $16^3 \times 24$
lattice at $\beta = 6.0$.
In the time direction,
fixed boundary conditions were imposed on the quarks to provide
larger time separations than available with periodic boundary
conditions. We also averaged over the directions of equivalent
lattice momenta in each configuration; this has the desirable effect of
reducing error bars.
Numerical details of this procedure are given in Refs. \cite{dwl90,ldd9495}
The dimensionless nucleon masses $M_N a$ for
$\kappa = 0.154$, 0.152, and 0.148 are 0.738(16), 0.882(12), and
1.15(1) respectively. The corresponding dimensionless
pion masses $m_{\pi} a$ are
0.376(6), 0.486(5), and 0.679(4). Extrapolating the nucleon and
pion masses to the chiral limit we determine $\kappa_c = 0.1567(1)$
and $m_N a = 0.547(14)$.
Using the nucleon mass to set the scale to study
nucleon properties~\cite{dll9596,ldd9495},
the lattice spacing $a^{-1} = 1.72(4)$ GeV is determined. The
three $\kappa's$ then correspond to quark masses of about 120, 200,
and 360 MeV respectively.
 
The strangeness current $\bar{s}\gamma_{\mu}s$ contribution appears in the
DI only.  The full details of the extraction can be found in Ref.~\cite{dlw}
and we satisfy ouselves here by simply quoting the results.
We use a monopole form to extrapolate $G_M^s(q^2)$ with nonzero
$q^2$ to $G_M^s(0)$, giving  $G_M^s(0) = - 0.36 \pm 0.20 $.  Correlations
are taken into account and the error is from jackknifing the fitting
procedure.  
A similar analysis is done for the strange Sachs electric form factor
$G_E^s(q^2)$ and we find that $G_E^s(0)$ is
consistent with zero as it should be.  We find for the electric mean-square
radius $\langle r_s^2 \rangle_E = -6 \frac{dG_E^s(q^2)}{dq^2}|_{q^2 = 0} =
- 0.061 \pm 0.003\,{\rm fm}^2$. 

\section*{Acknowledgments} 

Financial support from the Australian Research
Council is gratefully acknowledged.

\bb{dlw} S.J.\ Dong, K.F.\ Liu, and A.G.\ Williams, hep-ph/9712483,
to appear in Phys.\ Rev. D.

\bb{DIS} J.\ Ashman {\it et al.}, Nucl.\ Phys.\ {\bf B328}, 1 (1989);
   B.\ Adeva  {\it et al.}, Phys.\ Lett. B {\bf 329}, 399 (1994);
   K.\ Abe  {\it et al.}, Phys.\ Rev.\ Lett.\ {\bf 74}, 346 (1995).
 
\bb{gls91} J.\ Gasser, H.\ Leutwyler, and M.E.\ Sainio, Phys.\ Lett.\
   B {\bf 253}, 252, 260 (1991).
 
\bb{cheng76} T.P.\ Cheng, Phys. Rev. {\bf D 13}, 2161 (1976);
              {\bf D 38}, 2869 (1988).
 
\bb{gl82} J.\ Gasser and H.\ Leutwyler, Phys. Rep. {\bf 87}, 77 (1982).

\bb{SAMPLE97} B.\ Mueller {\it et al.}, SAMPLE Collaboration, Phys.\
Rev.\ Lett.\ {\bf 78}, 3824 (1997).
 
\bb{lei96} D. B.\ Leinweber, Phys.\ Rev.\ {\bf D 53}, 5115 (1996).

\bb{hpm97} S. T.\ Hong, B. Y.\ Park, and D. P. Min, nucl-th/9706008.

\bb{gi97} P.\ Geiger and N.\ Isgur, Phys. Rev. {\bf D 55}, 299 (1997).

\bb{cbk96} C. V.\ Christov {\it et. al.}, Prog. Part. Nucl. Phys.
{\bf 37}, 1 (1996).

\bb{pve91} Thomas Jefferson Lab. proposals \# PR-91-010, M. Finn and
P. Souder, spokespersons , \# PR-91-017, D. Beck, spokesperson,
and \# PR-91-004, E. Beise, spokesperson.

\bb{dll9596} S.J.\ Dong, J.-F.\ Laga\"e, and K.-F.\ Liu, Phys.\ Rev.\ Lett.
{\bf 75}, 2096 (1995); Phys.\ Rev.\ {bf D 54}, 5496 (1996).
 
\bb{dwl90} T.\ Draper. R. M.\ Woloshyn, and K. F.\ Liu, Phys.\ Lett.
{\bf 234}, 121 (1990); W.\ Wilcox, T.\ Draper, and K. F.\ Liu, 
Phys. Rev. {\bf D 46}, 1109 (1992).

\bb{ldd9495} K. F.\ Liu, S. J.\ Dong, T. Draper, J. M.\ Wu, and W.\ Wilcox,
Phys. Rev. {\bf D 49}, 4755 (1994);
K.F. Liu, S.J. Dong, T. Draper, and W. Wilcox, Phys. Rev. Lett.
{\bf 74}, 2172 (1995).
 

\vskip 1 cm
\thebibliography{References}

\bibitem{mandelstam} S.~Mandelstam, \rf{\pr}{D 20}{3223}{1979}

\bibitem{bp} N.~Brown and M.R.~Pennington, \rf{\pr}{D 39}{2723}{1989}

\bibitem{gribov} V.N.~Gribov, \rf{\np}{B 139}{19}{1978}; D.~Zwanziger,
\rf{\np}{B 378}{525}{1992}

\bibitem{stingl} M.~Stingl, \rf{\pr}{D 34}{3863}{1986}; 
\rf{\pr}{D 36}{651}{1987} 

\bibitem{bps} C.~Bernard, C.~Parrinello, A.~Soni,
\rf{\pr}{D49}{1585}{1994} 

\bibitem{mms} P.~Marenzoni, G.~Martinelli, N.~Stella, 
\rf{\np}{B 455}{339}{1995}; P.~Marenzoni {\em et al}, 
\rf{\pl}{B 318}{511}{1993}

\bibitem{letter} D.B.~Leinweber, J.I.~Skullerud,
A.G.~Williams, and C.~Parrinello\rf{\pr}{D 58}{031501}{1998}

\bibitem{bs} G.S.~Bali and K.~Schilling, \rf{\pr}{D 47}{661}{1993}

\bibitem{cornwall} J.~Cornwall, \rf{\pr}{D 26}{1453}{1982}

\bibitem{next} D.B.~Leinweber, C.~Parrinello,
J.I.~Skullerud, A.G.~Williams, in preparation.
\end{document}